# Design of deeply cooled ultra-low dissipation amplifier and measuring cell for quantum measurements with a microwave single-photon counter

O. G. Turutanov[1,2]*, A. M. Korolev[3], V. I. Shnyrkov[4,5], A. P. Shapovalov[4,5], M. Baránek[1], S. Kern[1], V. Yu. Lyakhno[2,4], P. Neilinger[1,6], M. Grajcar[1,6]

[1]Department of Experimental Physics, Comenius University,
 Mlynská dolina, 84248 Bratislava, Slovakia
[2]B.Verkin Institute for Low Temperature Physics and Engineering of NAS of Ukraine,
 47 Nauky ave., 61103 Kharkiv, Ukraine
[3]Institute of Radio Astronomy, NAS of Ukraine, 4 Mystetstv str., 61002 Kharkiv, Ukraine
[4]G.V. Kurdyumov Institute for Metal Physics of the NAS of Ukraine,
 36 Acad. Vernadsky Blvd., 03142 Kyiv, Ukraine
[5]Kyiv Academic University, 36 Vernadsky blvd., Kyiv 03142, Ukraine
[6]Institute of Physics, Slovak Academy of Sciences, Dúbravská cesta, Bratislava, Slovakia

*e-mail: oleh.turutanov@fmph.uniba.sk, turutanov@ilt.kharkov.ua

**Abstract**

The requirements and details of designing a measuring cell and low-back-action deeply-cooled amplifier for quantum measurements at 10 mK are discussed. This equipment is a part of a microwave single-photon counter based on a superconducting flux qubit. The high electron mobility transistors (HEMTs) in the amplifier operate in unsaturated microcurrent regime and dissipate only 1 microwatt of dc power per transistor. Simulated amplifier gain is 15 dB at 450 MHz with a high-impedance (~5 kΩ) signal source and standard 50-Ω output.

## 1. Introduction

During the last two decades, we have observed impressive progress in studying and developing superconducting quantum circuits, which provide a hardware platform for manipulating individual microwave photons [1].

The traditional approach to measuring qubits is based on the amplification of a weak microwave probe tone with frequencies of several gigahertz. In order to isolate the fragile quantum states from the noise of a quantum-limited microwave amplifier, bulky magnetic nonreciprocal circuit components (cooled circulators) are used. This complicates the creation of a small integrated measurement system, which is needed e.g. for quantum error correction (QEC). This problem became the motivation for the development of an alternative approach to high-fidelity measurement of the qubit state using a built-in microwave single photon counter [2]. These microwave counters offer high performance and scalability to create an integrated measurement system. Their operation principle is based on the transient response of the measuring resonator excited by a weak microwave probe tone, which is dispersively coupled to the qubit and displays two characteristic frequencies. The resonator signal is detected by the Josephson Photomultiplier (JPM) [2,3], or, in terms of formalism [4], the "bright" and "dark" cavity pointer states of the resonator pointer.

A technique based on a microwave photon counter provides a binary result of a projective quantum measurement directly in the "cold" zone, at the 10-millikelvin stage of the dilution refrigerator, and it eliminates the need for nonreciprocal circuit components between the qubit and measurement circuitry. However, even in Noisy Intermediate Scale Quantum (NISQ) processors, the presence of 50–100 microwave channels of probe signals can complicate the practical implementation of the entire design, in particular, due to the cross-talk between tightly integrated resonators.

Note that in a double-well asymmetric qubit potential there are two classically distinguishable JPM states in terms of magnetic flux, which can be measured using an RF SQUID [5] operating in the non-hysteretic regime with low driving amplitudes [6]. The output signal from the SQUID resonator is fed to an amplifier cooled to 10–20 mK. The role of the amplifier in the scheme of the single-photon microwave counter proposed by us [7] is to isolate the measuring electronics from a quantum system and amplify the signal. In this paper, we discuss the construction of a special cryogenic amplifier with low power consumption and high input impedance to be used with an RF SQUID, and some details of the design of the measuring cell.

## 2. Single microwave photon detection

The idea of using photon-induced transitions in the discrete spectrum of an artificial quantum system ("artificial atom"), in particular, superconducting qubits, for detecting single microwave photons is currently generally accepted and has been experimentally demonstrated [8–10]. For such a counter based on a flux qubit, under the assumption of low dissipation, the stationary spectra of the qubit were calculated [5] from Eq. (1). The Hamiltonian on the left side of the equation is obtained from the classical expression for the energy of a single-contact quantum interferometer [11] by replacing the classical charge and flux variables with the corresponding quantum mechanical operators

$$\left\{-\frac{\hbar^2}{2C}\cdot\frac{\partial^2}{\partial\Phi^2}+\left[-\frac{I_c\Phi_0}{2\pi}\cos 2\pi\frac{\Phi}{\Phi_0}+\frac{(\Phi-\Phi_e)^2}{2L}\right]\right\}\Psi(\Phi)=E\cdot\Psi(\Phi) \qquad (1)$$

where $\hbar$ is Planck constant/$2\pi$, $\Phi_0 = 2.07\cdot 10^{-7}$ Wb is superconducting magnetic flux quantum, $C$ and $I_c$ are electrical capacitance and critical current of the Josephson junction, correspondingly, $L$ is the qubit loop inductance, $\Phi$ and $\Phi_e$ are magnetic flux inside the loop and external magnetic flux, correspondingly, $\Psi$ is wave function and $E$ is the energy.

In the same work [5], the possibility of tuning the resonant frequency in a wide range (3–12 GHz) by changing the qubit parameters was shown.

Microwave photon-induced transitions between operating levels in upper well, followed by tunneling and relaxation to the lower levels of deeper well of the qubit's double-well potential, lead to a measurable change in the magnetic flux in the qubit ($\sim 0.3\,\Phi_0$), after which the counter is again set in the waiting state on the lower of the two working levels (Fig. 1). For this operating cycle, the qubit must have certain relaxation rate associated with both the internal processes and the influence of measurement circuits, which can be taken into account by the equivalent resistance of the Josephson junction $R_N \sim 10\,\text{k}\Omega$ [7], which lead to the equivalent quality factor of the qubit $Q \sim 100$ if we consider the qubit as an oscillator.

These two classical states, clearly distinguishable by the magnetic flux in the qubit loop, can be measured by an RF SQUID inductively coupled to the qubit and operating in the adiabatic non-hysteretic mode ($\beta_L = 2\pi I_c L/\Phi_0 = (0.85-0.9) < 1$) [6] using the procedure of weak continuous quantum measurements [12] (Fig. 2).

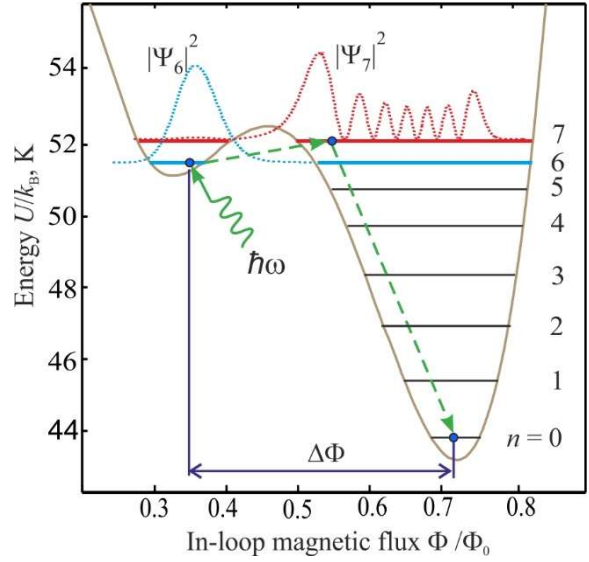

*Fig. 1.* Potential energy $U/k_B$ of the flux qubit in temperature units vs. magnetic flux inside the qubit loop $\Phi$ and its stationary spectrum. Squared wave function modules $|\Psi|^2$ for two operation levels 6 and 7 are shown. The arrows indicate the transitions upon absorption of a photon, which lead to a change in the flux by $\Delta\Phi$. Calculation parameters are $L = 2\cdot 10^{-10}$ H, $\beta_L = 1.325$, $C = 76$ fF, $\Phi_e = 0.5135\,\Phi_0$.

The qubit parameters during the working cycle are controlled by two magnetic gates, one of which governs the critical current of the Josephson junction, which is made as a "double rf SQUID" [13], and the other determines the external magnetic flux, i.e. the height of the barrier between the double-well potential wells, and the skew of the potential, respectively (see Fig. 2).

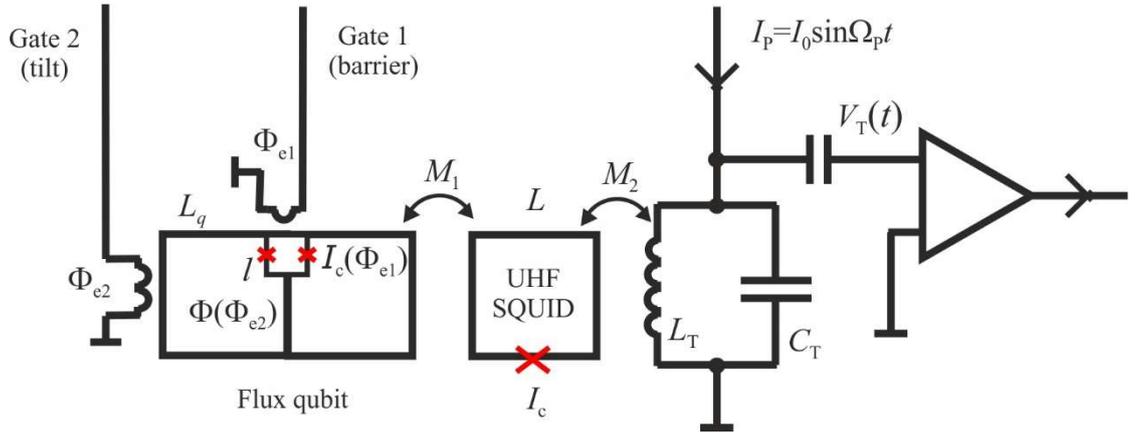

*Fig. 2.* Scheme of the weak continuous quantum measurements of a qubit: detection of the magnetic flux in the qubit loop by an inductively coupled RF SQUID. The qubit parameters during operation cycle are controlled by magnetic gates.

The advantages of an RF SQUID in the non-hysteretic mode are that (i) the Josephson junction does not go into resistive state and, therefore, does not generate noise that could cause qubit decoherence; (ii) small coupling coefficient with a qubit (~0.01) and (iii) small pump amplitude ( $\sim 0.001\,\Phi_0$) also reduce the backaction on the qubit counter. Thus, the RF SQUID operates as an almost ideal parametric upward converter. To increase the conversion factor, the pump frequency was increased up to 450 MHz. A complete block diagram of the counter is given in [7].

### 3. Isolation from electromagnetic environment and measuring cell

The isolation of a quantum system from external noisy environment which destroys the coherence of its quantum states is a major problem in quantum measurements [14]. For the microwave single-photon counter, it is also important to get low dark count rate. The solution includes thermalization the qubit and the measuring cell down to 10-mK stage of dilution refrigerator to block the Planck radiation; using combined superconducting and ferromagnetic magnetic shields; dielectric sample holder with high thermal conductivity to cool down the sample while avoiding thermal Nyquist currents and the corresponding noise magnetic fields; and filtering measuring circuits with wide-band powder microwave low-pass filters. Other special measures are the utilization of RF SQUID in non-hysteretic adiabatic regime with small driving amplitude and elevated pump frequency [6] and designing a cooled amplifier with low input brightness temperature to reduce the backaction (see the section 4 below).

To isolate a chip with the qubit, SQUID, and the preamplifier from slowly fluctuating magnetic fields, a hybrid shield was developed [15] consisting of two superconducting lead cylinders with a bottom (outer and inner) and a ferromagnetic shield made of Cryoperm G10 between them (Fig. 3(a)).

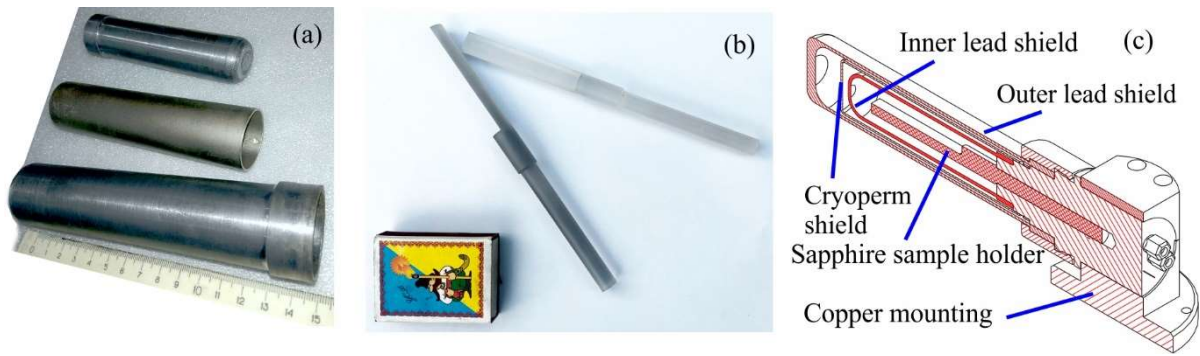

*Fig. 3.* (a) Two superconducting lead and one ferromagnetic Cryoperm G10 shields; (b) sapphire rod (sample holder). Match box (50 mm) is shown for scale; (c) scheme of the assembled measuring cell which is to be cooled down to 10 mK.

The calculated shielding coefficients are 200 dB for static magnetic field, and 70 dB for its low-frequency variations [15], i.e. the earth's magnetic field and industrial interference are effectively suppressed. The suppression of electrical noise in the measuring circuits is provided by chains of RC low-pass filters (above 500 MHz) and broadband powder filters similar to [16] installed at different temperature levels. To eliminate the effect of eddy Nyquist currents on the SQUID, the rod-shaped sample holder, which serves to cool it, is fabricated from sapphire (Fig. 3(b)). Although sapphire like other cryogenic materials dramatically loses its thermal conductivity as ($\sim T^3$) below liquid helium temperature [17] but it is still the best dielectric material among them at 10 mK. It has low dielectric losses at microwave frequencies which decrease with temperature [18]. The rod diameter is 8 mm, the length is 140 mm, with flat $10 \times 55$ mm area at its end. The lithographic chip with the qubit and RF SQUID, the SQUID's resonant circuit, and the preamplifier are placed on this pad. The entire measuring cell (Fig. 3(c)) is mounted on the bottom stage of a dilution refrigerator at 10 mK temperature. Note that the background photon flux at the operating frequency of the counter strongly depends on the temperature of the cooling stage; therefore, good thermal contact with the cooling plate and low heat generation by the amplifier inside the cell should be guaranteed. Using the Planck formula (2) for spectral radiance $B_f$ as a function of frequency $f$ and temperature $T$

$$B_f(f,T) = \frac{2hf^3}{c^2} \cdot \frac{1}{e^{hf/(k_B T)} - 1} \quad (2)$$

we can calculate spectral density of the electromagnetic power emitted inside the shields cavity and thus the rate of generation of thermal photons at different frequencies, close to the counter operation frequency as a function of the wall temperature. The results are shown in Table 1. The radiating area is $\sim 1.25 \cdot 10^{-2}$ m$^2$, which is the sum of inner shield cylinder and the sample holder surface areas.

**Table 1. Number of thermal photons at specified frequencies inside the magnetic shield cylinder**

| Frequency/Temperature | 8 GHz | 10 GHz | 12 GHz | 15 GHz |
| --- | --- | --- | --- | --- |
| 10 mK | 1 photon/14 days | 1 photon/200 years | - | - |
| 20 mK | 252 photon/s | 3.6 photon/s | 4 photon/min | 0.3 photon/h |
| 30 mK | 156000 photon/s | 11600 photon/s | 920 photon/s | 15 photon/s |

### 4. Cooled HEMT amplifier with ultra-low dissipation

The isolation of the qubit from the measuring circuits can be improved by using an amplifier based on HEMT-type transistors operating in the unsaturated mode [19]. Its main purpose is to match the impedances of the signal source and subsequent amplification stages and minimize a backaction. It was shown [20] that the brightness temperature $T_b$ of the input of the HEMT in the unsaturated regime considered as a black body practically coincides with the HEMT physical temperature, or the crystal lattice temperature $T_{latt}$. It is $T_b$ that characterizes the radiation of the amplifier towards the signal source (quantum system) and plays the main role in assessing the backaction upon it, which causes decoherence.

The brightness temperature should not be confused with the noise temperature. It corresponds to the physical temperature of the transistor and determines the thermal (Planck's) radiation directed to the signal source (a qubit). The noise temperature and equivalent noise power describe stochastic processes in solids governed by various mechanisms. Contrary to equivalent noise temperature, the brightness temperature does not follow the Friis theorem [21].

With good heat removal from the crystal, we can assume the lattice temperature $T_{latt}$ to be close to the ambient temperature $T_{amb}$. In major practical implementations of quantum measurements, cooled amplifiers are placed at 0.3-1 K temperature stage, since the cooling capacity of a dilution refrigerator at 10 mK is too small for amplifiers operating in the traditional saturated mode (usually about 100 μW). At the same time, in the saturated mode, HEMTs have an effective electron temperature of the 2D electron gas in the sink region $T_d$, which is two orders of magnitude higher than the physical temperature, contributing to the brightness temperature of the transistor input, according to the model [22]. Thus, for an ideal matching of complex impedances of a source and the amplifier:

$$T_b = T_g + |S_{12}| T_d \quad (3)$$

where $T_g$ is the gate temperature, and absolute value of S-parameter $|S_{12}|$ characterizes the portion of radiation travelling in backward direction.

The effect of $T_d$ for commonly used saturated HEMT regime can be roughly estimated assuming (see above) $T_g \approx T_{latt} \approx T_{amb}$, $T_d \approx 100\, T_{latt} \approx 100\, T_{amb}$ [22,23]. For the ultra-low-consumption (unsaturated) HEMT regime [19], as it was shown in [20], $T_d$ can be much smaller, down to $T_d \approx T_{latt} \approx T_{amb}$.

Typically, $|S_{12}|$ is about -20...-30 dB at 1 GHz frequency. In some cases the effect of $T_d$ can prevail over that of $T_g$.

A distinctive requirement for the development of an amplifier for a single-photon counter [7] is the high impedance of the signal source at a relatively high frequency (in contrast to a low-frequency cooled amplifier [24]). The signal source in the single-photon counter circuit is the resonant tank of the RF SQUID, which has a characteristic impedance of up to 5-6 kΩ at a resonant frequency of 450 MHz (quality factor $Q \sim 100$ and tank coil inductance $L_T \simeq 2 \cdot 10^{-9}$ H). At the same frequency, the input impedance of the transistor is about 600 Ω, so impedance matching circuit is required. To avoid thermal radiation from the coaxial cable, the amplifier should be placed in the vicinity of the SQUID tank circuit, i.e. at the 10-mK refrigerator stage, and this determines the maximum allowable power dissipation of the amplifier to be less than 10 microwatts. The amplifier output also should be matched to a standard 50-Ω coax line to carry the signal to next amplification stage. In this case, the amplifier must provide of at least 10 dB gain to reduce the influence of subsequent, warmer, cascades according to the Friis theorem [21].

The transfer of the amplifier from 1 K to 10 mK temperature level reduces its brightness temperature correspondingly, and its integral wide-band Planck radiation by 8 orders of magnitude. The narrow-band radiation at the counter operating frequency (10 GHz) per 1 Hz falls by unbelievable 21 orders of magnitude while at SQUID pump frequency (450 MHz) by 2 orders of magnitude, which is also good. Such a decrease in the brightness temperature is facilitated by small heat dissipation in the transistor and good heat removal from the crystal, so that its temperature is almost the same as that of lowest stage of the refrigerator. Meeting these conditions, including the gain, provides an extremely small backaction of the amplifier.

First we tested a wideband single-stage amplifier according to [19] in the form of a simulation and "in hardware", having measured it at temperatures of 300, 77 and 0.3 K, and assured that, with a high-impedance signal source, its gain was too low in unsaturated regime.

The next step was to design a two-stage-cooled narrow-band amplifier with matching circuits to achieve the specified requirements. Fig. 4 displays the wire diagram of the amplifier.

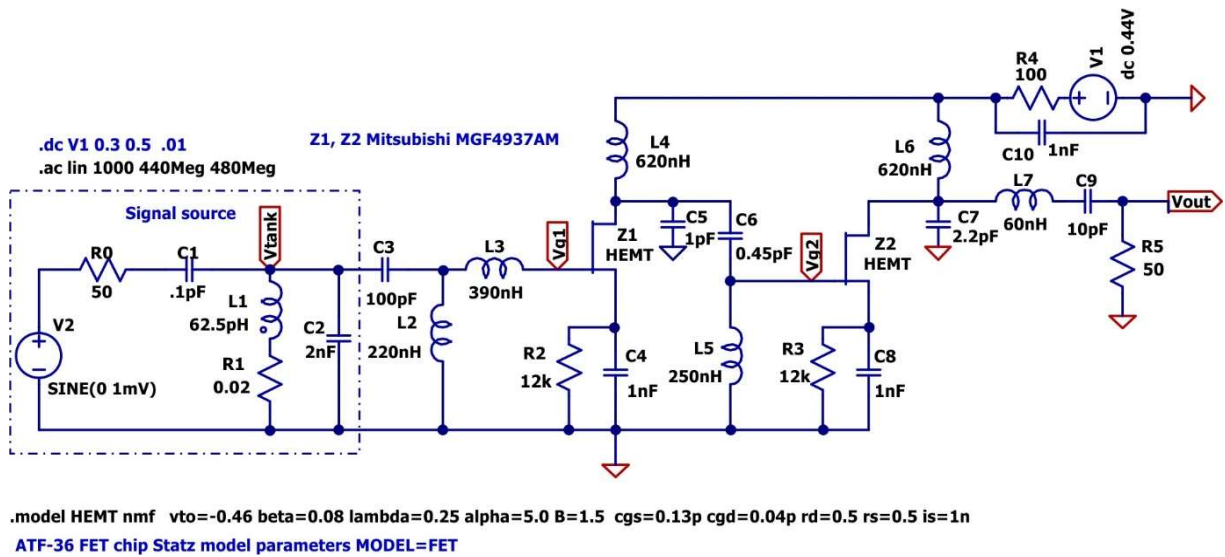

*Fig. 4*. Two-stage HEMT amplifier wire diagram for simulation in LTSPICE.

The Mitsubishi MGF4937AM was chosen instead of AVAGO ATF 36067 used in [19] since the latter one was already discontinued.

In the diagram, the dashed line box outlines the SQUID tank circuit with a specified quality factor and resonant frequency fed by an ac source V2. L2 and L3 compose an impedance matching circuit. C7 and L7 elements at the amplifier output have the same function. The dc supply voltage source V1 and gate auto bias resistors R2 and R3 determine the transistors operating point(s).

When we started designing the amplifier we faced the fact that the vendor gives no data for the low temperatures and low voltage/current in unsaturated regime. Therefore, we measured the static current-voltage characteristics of the transistor in the required range of voltages and currents corresponding to the unsaturated regime at the temperature of 0.35 K in a He-3 sorption refrigerator (Fig. 5).

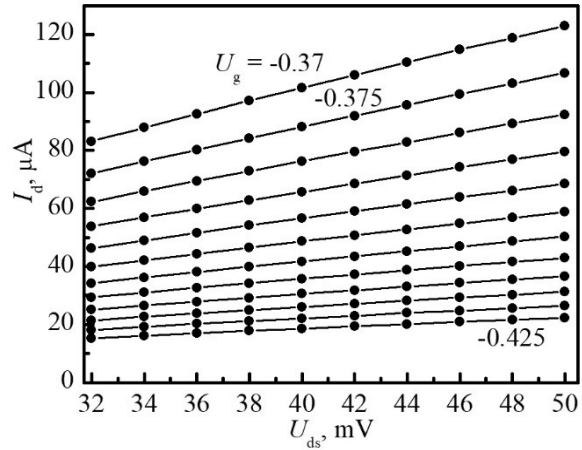

*Fig. 5.* The drain current $I_d$ vs. drain-source voltage $U_{ds}$ for the Mitsubishi MGF4937AM transistor in its unsaturated mode, measured at various gate bias $U_g$ at a temperature of 0.35 K.

The output resistance $R_d$, transconductance $g_{dg}$ and power $P_{HEMT}$ dissipated in the HEMT channel were derived from these data.

To simulate the circuitry in LTSPICE program, we also had to correct the standard coefficients in the Statz model [25] of GaAs FETs built in LTSPICE by fitting the calculated and the experimental curves. Only two coefficients needed correction, namely, the transconductance parameter $\beta$, from 0.1 to 0.08, and the threshold gate voltage $U_T$, from -0.55 to - 0.46 (Equation (4)). According to [25], the drain current is

$$I_d(U_{gs}, U_{ds}) = \beta(U_{gs} - U_T)^2 (1 + \lambda U_{ds}) \tanh(\alpha U_{ds}), \qquad (4)$$

where $\beta$ is the transconductance parameter, $U_T$ is threshold voltage, $\lambda$ is channel length modulation parameter, $\alpha$ is hyperbolic tangent function parameter.

The equivalent input capacitance and resistance of the transistor, which change slightly with temperature, were calculated from the manufacturer's S-parameter table. This enabled the calculation of the matching circuit between the high-impedance resonant tank and the low-impedance input of the transistor.

After dc and ac simulations, we determined the required supply voltages, offsets, and parameters of matching resonant circuits (Fig. 6).

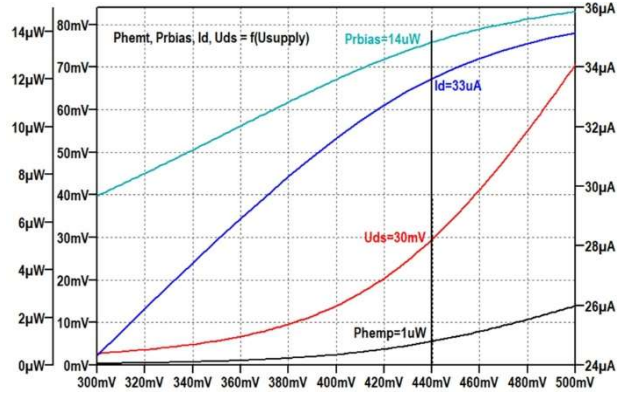

*Fig. 6.* Drain current $I_d$, drain-source voltage $U_{ds}$, transistor-dissipated power $P_{HEMT}$ and power dissipated by the gate-bias resistor $P_{bias}$ as functions of supply voltage $U_{supply}$ (dc simulation in LTSPICE). Vertical line and shown values indicate the chosen operation point.

We choose the operation point at $U_{ds} = 30$ mV, $I_d = 33.6$ μA corresponding to supply voltage 0.44 V by the criterion of the HEMT-dissipated power $P_{HEMT} = 1$ μW.

To provide ultra-low dissipation of the amplifier of a few microwatt, the automatic gate bias resistors R2 and R3 (Fig. 4) which dissipate an order of magnitude higher power (13.6 μW) than HEMT itself (1 μW), should be moved to a higher temperature level where the refrigerator's cooling capacity is larger, and connected via twisted pairs.

Simulation shows (Fig. 7) that there is no voltage gain in this unsaturated regime ($G_V = -6$ dB) but the current gain $G_I$ is 21 dB; in total, power gain $G_P$ is 15 dB in a narrow frequency band needed for better signal/noise ratio.

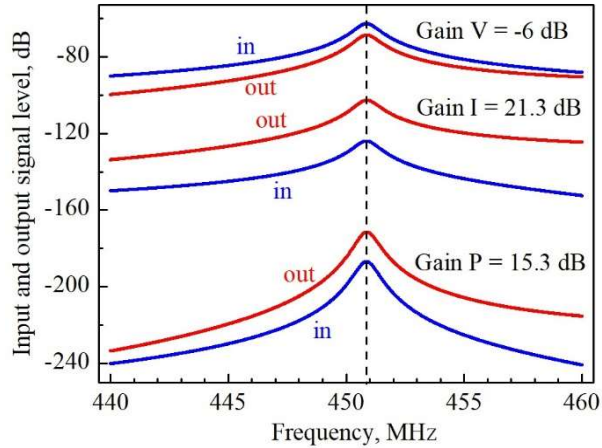

*Fig. 7.* The ac signal voltage (*V*), current (*I*) and power (*P*) vs. frequency at input (*in*) and output (*out*) of the cooled two-stage amplifier at operating point $U_{ds} = 30$ mV, $I_d = 30$ μA and temperature 0.35 K (ac simulation in LTSPICE). Corresponding gains are shown.

## 5. Conclusions

The development of a quantum measurement system, including a single microwave photon counter, is a complex task in which the main role is played by the protection (isolation) of the quantum system from the influence of the external electromagnetic environment and the backaction of the measuring equipment. In addition to cooling, shielding and filtering electrical circuits, it is extremely important to choose the right measurement technique and provide low backaction ensured by the special design of the first cooled amplifier stage. The described qubit isolation measures satisfy these requirements. Circuit simulation and low-temperature measurements of the transistor show that the developed two-stage cooled amplifier based on HEMT operating in unsaturated mode has an ultra-low power dissipation (2 μW), low backaction since the brightness temperature of its input is almost equal to the temperature of the lowest cooling stage (10 mK), and the amplifier itself has a fairly large gain (15 dB). Its feature is that its input matches the high-impedance signal source (the RF SQUID resonant tank) at a rather high-frequency (450 MHz). The amplifier output is also matched to a standard $50-\Omega$ transmission line for further signal processing. Such an amplifier can also be used in other experiments where the high-frequency signal source has a high impedance, and low backaction is required to measure a quantum object.

**Acknowledgments**

The work was supported by SPS Programme NATO grant number G5796 and funded by the EU NextGenerationEU through the Recovery and Resilience Plan for Slovakia under the project No. 09I03-03-V01-00031.

**References**

1. X. Gu, A. F. Kockum, A. Miranowicz, Y.-X. Liu, F. Nori, Phys. Rep. 718-719, 1 (2017). https://doi.org/10.1016/j.physrep.2017.10.002

2. A. Opremcak, I. V. Pechenezhskiy, C. Howington, B. G. Christensen, M. A. Beck, E. Leonard, J. Suttle, C. Wilen, K. N. Nesterov, G. J. Ribeill, T. Thorbeck, F. Schlenker, M. G. Vavilov, B. L. T. Plourde, and R. McDermott, Science **361**, 1239 (2018). https://doi.org/10.1126/science.aat4625

3. A. Opremcak, C. H. Liu, C. Wilen, K. Okubo, B. G. Christensen, D. Sank, T. C. White, A. Vainsencher, M. Giustina, A. Megrant, B. Burkett, B. L. T. Plourde, and R. McDermott, Phys. Rev. X **11**, 011027 (2020). https://doi.org/10.1103/PhysRevX.11.011027

4. W. H. Zurek, Phys. Rev. D **24**, 1516 (1981) https://doi.org/10.1103/PhysRevD.24.1516

5. V. I. Shnyrkov, Wu Yangcao, A. A. Soroka, O. G. Turutanov, V. Yu. Lyakhno, Fiz. Nizk. Temp. **44**, 281 (2018) [Low Temp. Phys. **44**, 213 (2018)]. https://doi.org/10.1063/1.5024538

6. V. I. Shnyrkov, A. P. Shapovalov, V. Yu. Lyakhno, A. O. Dumik, A. A. Kalenyuk and P. Febvre, Supercond. Sci. Technol. **36,** 035005 (9pp) (2023). https://doi.org/10.1088/1361-6668/acb10e

7. V. I. Shnyrkov, W. Yangcao, O. G. Turutanov, V. Y. Lyakhno and A. A. Soroka, 2020 IEEE Ukrainian Microwave Week (UkrMW), Kharkiv, Ukraine, 2020, pp. 737-742, https://doi.org/10.1109/UkrMW49653.2020.9252799

8. Y.-F. Chen, D. Hover, S. Sendelbach, L. Maurer, S. T. Merkel, E. J. Pritchett, F. K. Wilhelm, R. McDermott, Phys. Rev. Lett. **107**, 217401 (2011). https://doi.org/10.1103/PhysRevLett.107.217401

9. K. Inomata, Zh. Lin, K. Koshino, W.D. Oliver, J.-S. Tsai, Ts. Yamamoto, and Ya. Nakamura, Nat. Commun. **7**, 12303 (2016). https://doi.org/10.1038/ncomms12303

10. R. Lescanne, S. Deléglise, E. Albertinale, U. Réglade, T. Capelle, E. Ivanov, T. Jacqmin, Z. Leghtas, and E. Flurin. Phys. Rev. X **10**, 021038 (2020). https://doi.org/10.1103/PhysRevX.10.021038

11. A. Barone and G. Paternò, Physics and Applications of the Josephson Effect (Wiley, New York, 1982).

12. A. N. Korotkov and D. V. Averin, Phys. Rev. B **64**, 165310 (2001). https://doi.org/10.1103/PhysRevB.64.165310

13. S. Han, J. Lapointe, J. E. Lukens. Variable ß RF SQUID. In: H. Koch, H. Lübbig, (eds.) Single-Electron Tunneling and Mesoscopic Devices. Springer Series in Electronics and Photonics, vol. 31. Springer, Berlin, Heidelberg, 1992, p.219. https://doi.org/10.1007/978-3-642-77274-0_25

14. V. I. Shnyrkov, A. M. Korolev, O. G. Turutanov, V. M. Shulga, V. Yu. Lyakhno, V. V. Serebrovsky, Fiz. Nizk. Temp. **41**, 1109 (2015) [Low Temp. Phys. **41**, 867 (2015)]. https://doi.org/10.1063/1.4935839

15. V. Yu. Lyakhno, O. G. Turutanov, A. P. Boichenko, A. P. Shapovalov, A. A. Kalenyuk, V. I. Shnyrkov, Fiz. Nizk. Temp. **48**, 254 (2022) [Low Temp. Phys. **48**, 228 (2022)]. https://doi.org/10.1063/10.0009541

16. J. M. Martinis, M. H. Devoret, J. Clarke, Phys. Rev. B **35** 4682 (1987). https://doi.org/10.1103/PhysRevB.35.4682

17. R. Berman. Proc. R. Soc. London **A 208**, 90 (1951). https://doi.org/10.1098/rspa.1951.0146

18. J. Krupka, K. Derzakowski, M. Tobar, J. Hartnett and R. G Geyer, Meas. Sci. Technol. **10**, 387 (1999). https://doi.org/10.1088/0957-0233/10/5/308

19. A.M. Korolev, V.I. Shnyrkov, and V.M. Shulga, Rev. Sci. Instrum. **82**, 016101 (2011), https://doi.org/10.1063/1.3518974

20. A.M. Korolev, V.M. Shulga, O.G. Turutanov, V.I. Shnyrkov, Solid State Electron. **121**, 20 (2016). https://doi.org/10.1016/j.sse.2016.03.010

21. H.T. Friis. Noise figures of radio receivers. Proc. IRE **32**(7), 419 (1944). http://doi.org/10.1109/JRPROC.1944.232049

22. M. W. Pospieszalski. Modeling of noise parameters of MESFET's and MODFET's and their frequency and temperature dependence. IEEE Trans Microw Theory Tech **37**(9), 1340 (1989). http://dx.doi.org/10.1109/22.32217

23. J. Stenarson, M. Garcia, I. Angelov, H. Zirath. A general parameter-extraction method for transistor noise models. IEEE Trans Microw Theory Tech **47**(12), 2358 (1999). http://dx.doi.org/10.1109/22.808982

24. A.M. Korolev, V.M. Shulga, I.A. Gritsenko, G.A. Sheshin, Cryogenics **67**, 31 (2015). https://doi.org/10.1016/j.cryogenics.2015.01.003

25. H. Statz, P. Newman, I. W. Smith, R. A. Pucel, H. A. Haus, IEEE Trans. Electron Devices **34**, 160 (1987). https://doi.org/10.1109/T-ED.1987.22902.